\documentclass[aps,prl,twocolumn,floatfix]{revtex4}

\usepackage{graphicx}

\def\gtorder{\mathrel{\raise.3ex\hbox{$>$}\mkern-14mu
 \lower0.6ex\hbox{$\sim$}}}
\def\ltorder{\mathrel{\raise.3ex\hbox{$<$}\mkern-14mu
 \lower0.6ex\hbox{$\sim$}}}

\def\gep{G_E}

\begin{document}

\title{Comment on ``Breakdown of the expansion of finite-size corrections to
the hydrogen Lamb shift in moments of charge distribution''}

\author{J. Arrington}

\affiliation{Physics Division, Argonne National Laboratory, Argonne, Illinois 60439, USA}

\begin{abstract}

In a recent paper, Hagelstein and Pascalutsa examine the error associated with
an expansion of proton structure corrections to the Lamb shift in terms of
moments of the charge distribution. They propose a small modification to a
conventional parameterization of the proton's charge form factor and show that
this can resolve the proton radius puzzle. However, while the size of the
"bump" they add to the form factor is small, it is large compared to the total
proton structure effects in the initial parameterization, yielding a final form
factor that is unphysical. Reducing their modification to the point where the
resulting form factor is physical does not allow for a resolution of the
radius puzzle.

\end{abstract}

\date{\today}

\maketitle


Ref.~\cite{hagelstein15} proposes a possible explanation to the proton radius
puzzle~\cite{pohl13, carlson15}, noting that the error associated with the
expansion of the Lamb shift in terms of the moments of the charge radius,
$\langle r^2 \rangle$ and $\langle r^3 \rangle$, can be large in the presence
of sharp structures in the form factors. They demonstrate that a small,
narrow contribution to the proton's charge form factor at very low-$Q^2$ could
explain the discrepancy in the extracted RMS charge radius from the muonic
hydrogen Lamb shift measurements~\cite{pohl10, antognini13}. Their example
involved a narrow peak added to a standard parameterization of the charge
form factor, $\gep(Q^2)$, at $Q^2$ values which significantly impact the Lamb
shift in muonic hydrogen~\cite{pohl10, antognini13}. The modification is too
high in $Q^2$ to significantly modify the Lamb shift in electronic
hydrogen~\cite{parthey11, mohr12}, but below the $Q^2$ region where electron
scattering data exist and can be used to extract the charge
radius~\cite{sick03, bernauer10, zhan11, lee15, arrington15a}.

\begin{figure}[tb]
\begin{center}
\includegraphics[width=0.40\textwidth]{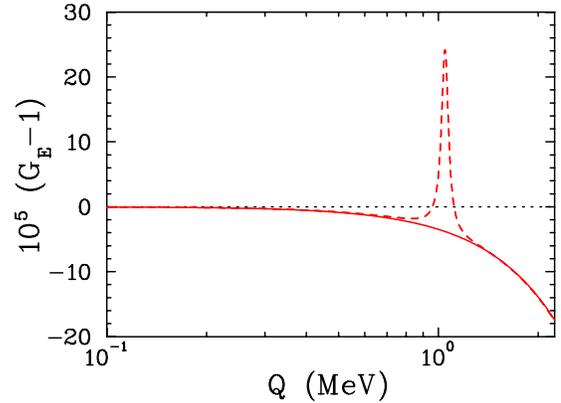}
\caption{The finite structure correction to the $\gep$,
$\gep(Q^2)-1$, vs $Q$ for the parameterization of Ref.~\cite{arrington07b}
(solid line), and including the modification of Ref.~\cite{hagelstein15}
(dashed line).}
\label{fig:fig1}
\end{center}
\end{figure}

Their proposed modification to $\gep$ is very small, with a peak contribution to
$\gep$ of $3 \times 10^{-5}$, narrowly localized around $Q^2 \approx
10^{-6}$~GeV$^2$. However, while the change in $\gep$ is extremely small, that
does not mean that this is a minor modification to the proton form factor.
This modification should not be compared to $\gep$, which is close to unity at
low $Q^2$, but should should be compared to $\gep-1$ which represents the
deviation of the form factor from that of a point proton: $\gep(Q^2)=1$. For
the form factor parameterization~\cite{arrington07b} used
in~\cite{hagelstein15}, $| \gep-1 | = 3.5 \times 10^{-6}$ for $Q^2 \approx
10^{-6}$~GeV$^2$. The proposed modification, while small compared to $\gep$,
is roughly ten times larger than the total finite-size effect
in~\cite{arrington07b}. Because this bump is added to the form factor, their
modified form factor is unphysical, yielding $\gep > 1$ as shown in
Figure~\ref{fig:fig1}.

Based on Fig. 3 of Ref.~\cite{hagelstein15}, reducing the size of the
modification by an order of magnitude to avoid $\gep>1$ would not provide a
significant improvement in the agreement between eH and $\mu$H Lamb shift
results. Similar features in the region of the eH sensitivity peak, $Q^2
\approx 10^{-10}$~GeV$^2$, would have to be 10$^5$ times smaller to avoid
exceeding the full finite-size correction from~\cite{arrington07b}. Even if a
smaller (or negative) modification were made, such that the resulting $\gep$
would not be unphysical, it would most likely be inconsistent with the
constraints from analyticity of the form factors~\cite{hill10}.

While the bump added to $\gep$ in~\cite{hagelstein15} brings the Lamb shift
extractions into agreement after correcting for the error made when expanding
in moments of the charge radius, the resulting form factor is unphysical.
Simply reducing or broadening the bump near the peak of sensitivity for the
$\mu$H Lamb shift measurements cannot provide a resolution to the discrepancy.
It seems unlikely that it is possible to find another such modification which
resolves the discrepancy and is consistent with the constraints from the
analyticity of the form factors.

This work was supported by the U.S. Department of Energy, Office of Science,
Office of Nuclear Physics, under contract DE-AC02-06CH11357.

\bibliography{comment_bump}

\end{document}